# S-parameter calibration procedure for multiport microwave imaging systems

Manuel Kasper[1], Mykolas Ragulskis[1], Georg Gramse[1,2], Ferry Kienberger[1]
[1] Keysight Laboratories, Keysight Technologies Inc., Gruberstrasse 40, 4020 Linz, Austria
[2] Johannes Kepler University, Biophysics, Gruberstrasse 40, 4020 Linz, Austria,

Correspondence manuel.kasper@keysight.com

*Abstract*—Multitude of antennas are typically used in microwave imaging systems. Here we outline a simple and effective calibration method for multiport imaging systems. By using only one additional component, an electronic calibration module (ECal), one port is calibrated and the calibration plane is thereby moved to the antenna connector. Assuming all antennas interact with the test phantom in the same way, the one-port calibration is transferred to all other antennas. For full calibration including the transmission path between antennas, the "unknown thru" technique is used. This calibration procedure is simple and it can be fully automated, and no RF components are perturbed.

*Index Terms*—microwave imaging, VNA calibration, multiport.

## I. Introduction

Microwave imaging systems typically have a multitude of antennas [1]. Multiport vector network analyzers (VNA) are used to excite the antennas and acquire measurement data. In many systems microwave switches are used to extend the number of available analyzer ports and to match the number of antennas [2,3]. To perform traditional multiport calibration [4,5], calibration standards must be connected and antennas replaced, thereby significantly increasing the amount of labor and system complexity. This work here presents a new calibration method where the calibration procedure can be fully automated and performed in-situ. The only user interaction required is to put the test-phantoms with known dielectric values into the system.

## II. Method

### A. System requirements

There are only few assumptions on the imaging system required for the new calibration. Firstly, all antennas are assumed to be the same, meaning that when measuring homogenous phantoms all antennas have the same reflection coefficient. This implies also that the distance between antennas and the phantom is constant for all antennas. Secondly, three electrically distinct homogenous phantoms are required. Air is a first choice, and two other test phantoms can be water and alcohol if the system can be filled with liquids. Alternatively, solid or gel phantoms can be used if the distance between antennas and phantom is matched. The knowledge of the exact permittivity of the phantom materials is thereby not necessary. Thirdly, one ECal module has to be installed on one of the antennas as shown in Fig. 1. The ECal module is thereby permanently kept in the imaging system at this specific antenna and it is not removed during the complete calibration procedure.

### B. Imaging system calibration procedure

The first step in the calibration procedure is to calibrate one of the ports for reflection. This is done by connecting an ECal directly to one selected antenna port. By performing a standard reflection calibration procedure, the calibration plane is defined at the ECal connector connected to the VNA port. To shift the calibration plane to the antenna connector (Fig. 1 dashed red line) the ECal is switched into its "thru" state and the "thru" is de-embedded. Since the "thru" state of the ECal is known its influence on all following measurements can be mathematically removed..

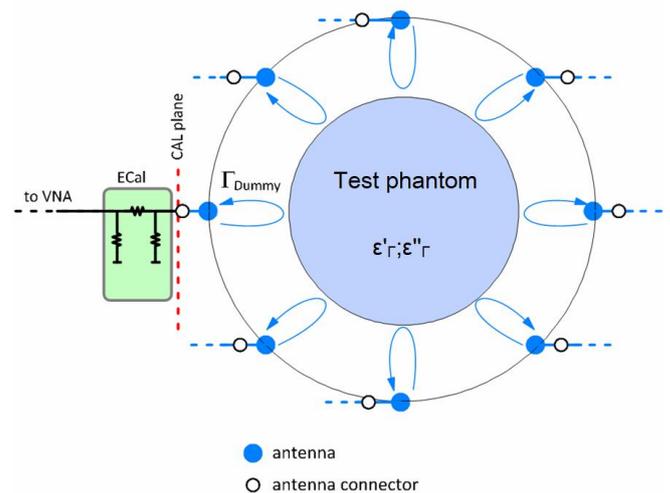

Fig. 1. Reflection calibration of a single port using the ECal and de-embedding at the calibration plane (dotted line). The interaction between antennas and the test phantom is illustrated with arrows.

The next step is to transfer the reflection calibration to all remaining antennas. Thereby, the port calibrated with the ECal is used to measure the reflection coefficients of three test phantoms (test phantom 1, 2, and air). The same measurement is performed with all other uncalibrated ports. The calibrated reflection measurements are then used as standards to calculate the error coefficients of the full error box (Fig. 2). This calibration step effectively shifts the calibration planes to the various antenna connectors.

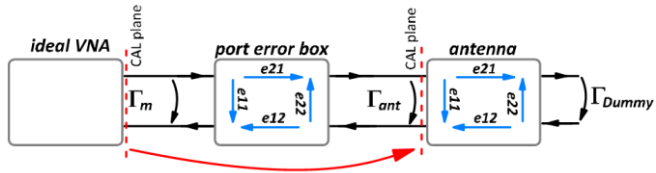

Fig. 2. Simplified signal flow graph showing the transfer of the reflection calibration from one port to the remaining ports. All antennas are assumed to have the same error coeffients.

In the final step and to achieve full calibration, it is necessary to calibrate each individual path between all the antennas for transmission. This is done by employing the "unknown thru" technique [6] where the coupling between the antennas act as "unknown thru" (see Fig 3). The only requirement here is that the transmission path is fully reciprocal, i.e. S12 = S21, which is true for passive devices. A technical requirement is also that enough signal is transmitted to avoid noise issues in the data acquisition. This is typically fulfilled if the path attenuation is not more than −50 dB which is normally the case for multiport imaging systems.

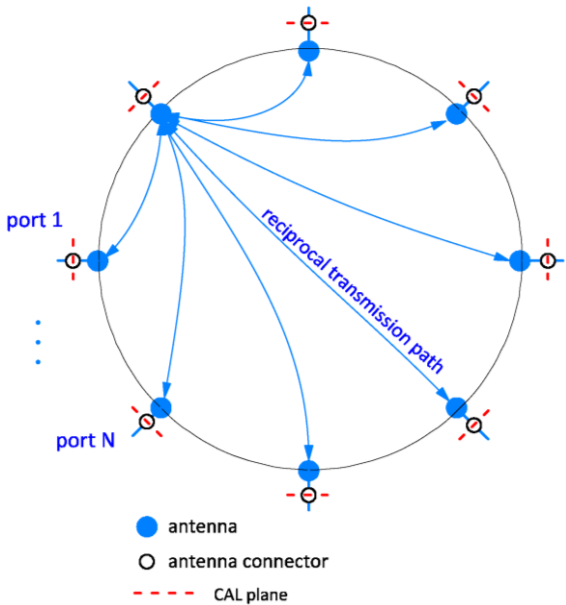

Fig. 3. Calibrating each path for transmission by use of the "unknown thru" technique. Through measurements are thereby performed for each antenna.

### III. DISCUSSION AND CONCLUSIONS

If multiport systems are measured by a set of two-port measurements some ports are inactive and therefore they need to be terminated. The termination of the inactive ports theoretically effects the measured S-parameters. Rautio [7] describes a method for correcting the effects of termination. While this is relevant for low-loss multiport devices, the path-loss between the antennas of imaging systems greatly attenuates this effect and therefore it can be neglected in most cases.

Although all the assumptions and requirements outlined above fit quite well with most microwave imaging systems, there are exceptions. For example, systems where antennas can be moved in relation to each other antenna, or non-symmetrical arrangements of test phantoms resulting in different interactions between antennas and phantoms on each antenna are such exceptions. Additionally, many system designers simplify microwave hardware by using a RF switching matrix to switch antennas between transmit and receive modes. In such a case, reflection measurements become impossible thus preventing one port calibration.

However, if the requirements are met, then by implementing the proposed method several advantages are achieved. This includes potential savings in costs for calibrating the system in production environment, and latter for recalibration at the factory. Also because of the simplicity of this method, the possibility to calibrate in short intervals (e.g. each day) increases the reliability of measured data [8].

### ACKNOWLEDGMENT

This work was supported by the European Union's Horizon 2020 research and innovation program under the Marie Skłodowska-Curie grant agreement No. 764479 (Emerald) and the project MMAMA under Grant Agreement Number 761036.